\documentclass[journal=jacsat,manuscript=article]{achemso}

\usepackage{graphicx}

\usepackage{physics}
\usepackage{amsmath}
\usepackage[eulergreek]{sansmath}
\usepackage{svg}
\usepackage{nameref}
\usepackage{varioref}
\usepackage{hyperref}
\hypersetup{
    colorlinks=true,
    linkcolor=blue,
    filecolor=magenta,      
    urlcolor=cyan,
    citecolor=blue,
}
\usepackage{cleveref}

\usepackage{graphicx}
\usepackage{xcolor}
\usepackage{subcaption}

\sansmath
\newcommand{\tens}[1]{\vb{\hat{#1}}}

\newcommand{\GE}[0]{\vb{\hat{G}}^{\rm E}}
\newcommand{\GH}[0]{\vb{\hat{G}}^{\rm H}}

\newcommand{\rGE}[0]{\curl \vb{\hat{G}}^{\rm E}}
\newcommand{\rGH}[0]{\curl \vb{\hat{G}}^{\rm H}}

\author{Semyon Borodulin}
\affiliation[ITMO University]
{Department of Physics and Engineering, ITMO University, Saint-Petersburg}

\author{Natalia Kostina}
\affiliation[ITMO University]
{Department of Physics and Engineering, ITMO University, Saint-Petersburg}

\author{Mihail Petrov}
\affiliation[ITMO University]
{Department of Physics and Engineering, ITMO University, Saint-Petersburg}

\email{m.petrov@metalab.ifmo.ru}

\title[Directional Scattering-Induced Optical Forces on a Mie Particle near a Metal Interface]
  {Directional Scattering-Induced Optical Forces on a Mie Particle near a Metal Interface}

\begin{document}

\maketitle

\begin{abstract}
Optical manipulation of Mie-resonant dielectric nanoparticles is strongly influenced by their enhanced scattering and multipolar response, which fundamentally modify the balance of optical forces. In this work, we study the optical forces acting on a resonant dielectric nanoparticle placed near a metal interface, where scattering occurs into both free-space and surface plasmon-polariton (SPP) channels. We show that the interference of electric and magnetic dipole moments leads to highly directional scattering in these channels, and the direction and magnitude of the scattering-induced force are directly linked to the angular directivity of the corresponding radiation channels. We show that in a cross-beam configuration, where the radiation-pressure contribution is suppressed, the optical force can be changed over almost 2$\pi$ in a wide range of particle sizes, providing a route toward optical sorting of resonant nanoparticles. 
    
\end{abstract}
\newpage
\section{INTRODUCTION}
\begin{figure}
    \centering
\includegraphics[width=0.5\linewidth]{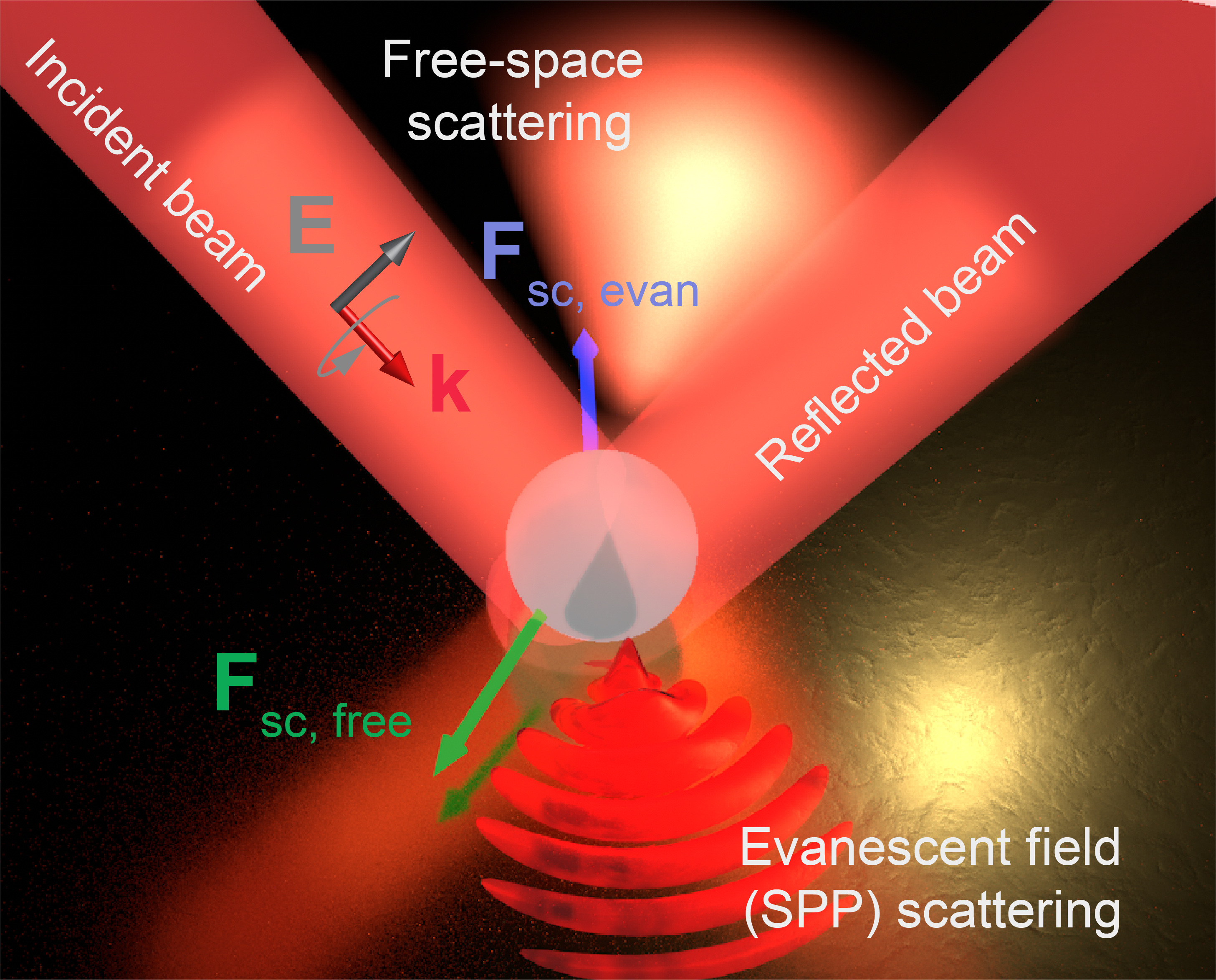}
    \caption{Directional scattering of a Mie-resonant nanoparticle near a metal interface into free-space and SPP channels generates a recoil optical force governed by momentum redistribution.}
    \label{fig:scheme}
\end{figure}

High-index dielectric nanoparticles supporting Mie resonances have become a central platform of modern nanophotonics. The coexistence of electric and magnetic multipolar modes in subwavelength particles enables low-loss control of scattering, near-field localization, and radiation directivity, underpinning applications in nonlinear optics, nanolasing, and biosensing \cite{Kuznetsov2016,Kivshar2022,Sain2019,Tseng2021}. Yet, compared with the rapid progress in these areas, the optical manipulation of Mie-resonant particles remains considerably less well developed. This limitation originates from the very nature of resonant scattering. Near Mie resonances, dielectric nanoparticles exhibit large scattering cross-sections, so that radiation pressure and recoil compete with the restoring gradient force \cite{stilgoe_effect_2008, nieminen_effect_2009, andres-arroyo_optical_2016, valuckas_fabrication_2019}. As a result, stable trapping in conventional single-beam optical tweezers becomes more demanding and often requires beam engineering or interference-based configurations \cite{AndresArroyo2016,Lank2018}. Recent studies further revealed that resonant particles can exhibit complex dynamics, including hopping between metastable positions and orbiting motion, even in simple trapping geometries \cite{toftul_hopping_2025}. In addition, the optical force can be actively tuned and even reversed in resonant systems, as demonstrated for phase-change nanoparticles with switchable attractive and repulsive regimes \cite{Mao2024Switchable}. Alternatively, the interplay between the attractive and repulsive interaction can lead to levitation regime of Mie particles near a dielectric waveguide structure \cite{shilkin_optical_2022} or form stable structure in the ensemble of resonant particles \cite{bulgakov_giant_2020, bulgakov_resonant_2021}. These results highlight the fundamentally multipolar and non-conservative nature of optical forces in the Mie regime.

From a practical perspective, an important unresolved problem is the sorting of resonant nanoparticles by size or spectral response. Conventional post-synthesis approaches, such as density-gradient separation of silicon colloids \cite{Karsakova2023}, can reduce size dispersion but do not enable in-situ, light-driven sorting. Optical forces offer an alternative route, where strong scattering can be exploited as a resource for force engineering. This concept underlies directional optical sorting and force control based on electric–magnetic interference \cite{Shilkin2017, Lank2018, Toftul2023PRL,Toftul2025ACS}.

Moreover, Mie-resonant nanoparticles exhibit highly directional scattering not only in free space \cite{Fu2013,Neugebauer2016, Kiselev2020MultipoleScattering}, but also in their coupling to surface plasmon-polariton (SPP) modes near metal interfaces \cite{Sinev2018,Sinev2021}. Surface-wave-assisted optical manipulation has been extensively studied for effectively dipolar particles, where SPP excitation can generate pulling, trapping, or anti-trapping forces \cite{Petrov2016,Ivinskaya2017, Kalhor2016, giron-sedas_strong_2020}. While the  plasmon assisted optical  manipulation methods have been actively developed \cite{Juan2011, Zhang2021}, the  optical force acting on resonant Mie particles near the dielectric and plasmonic interfaces has not been widely studied.  

In this paper, we extend the directional scattering of SPPs to the regime of multipolar interference in resonant dielectric particles, where the interplay of electric and magnetic modes enables additional control over scattering channels and optical forces (see \ref{fig:scheme}). We show that two dominant scattering channels — free space scattering and scattering into the evanescent (SPP) modes — contribute almost independently to the recoil   optical force. The sign and amplitude of these forces strongly vary with the particle diameter, due to the spectral shift of Mie resonances. Based on this, we propose the cross-beam geometry with almost $2\pi$ control of the direction  of lateral optical force, which opens a route to the optical sorting of Mie particles due to the directional scattering into the   free-space  and SPP modes.

\section{RESULTS AND DISCUSSION}

\begin{figure}
    \centering
    \includegraphics[width=\linewidth]{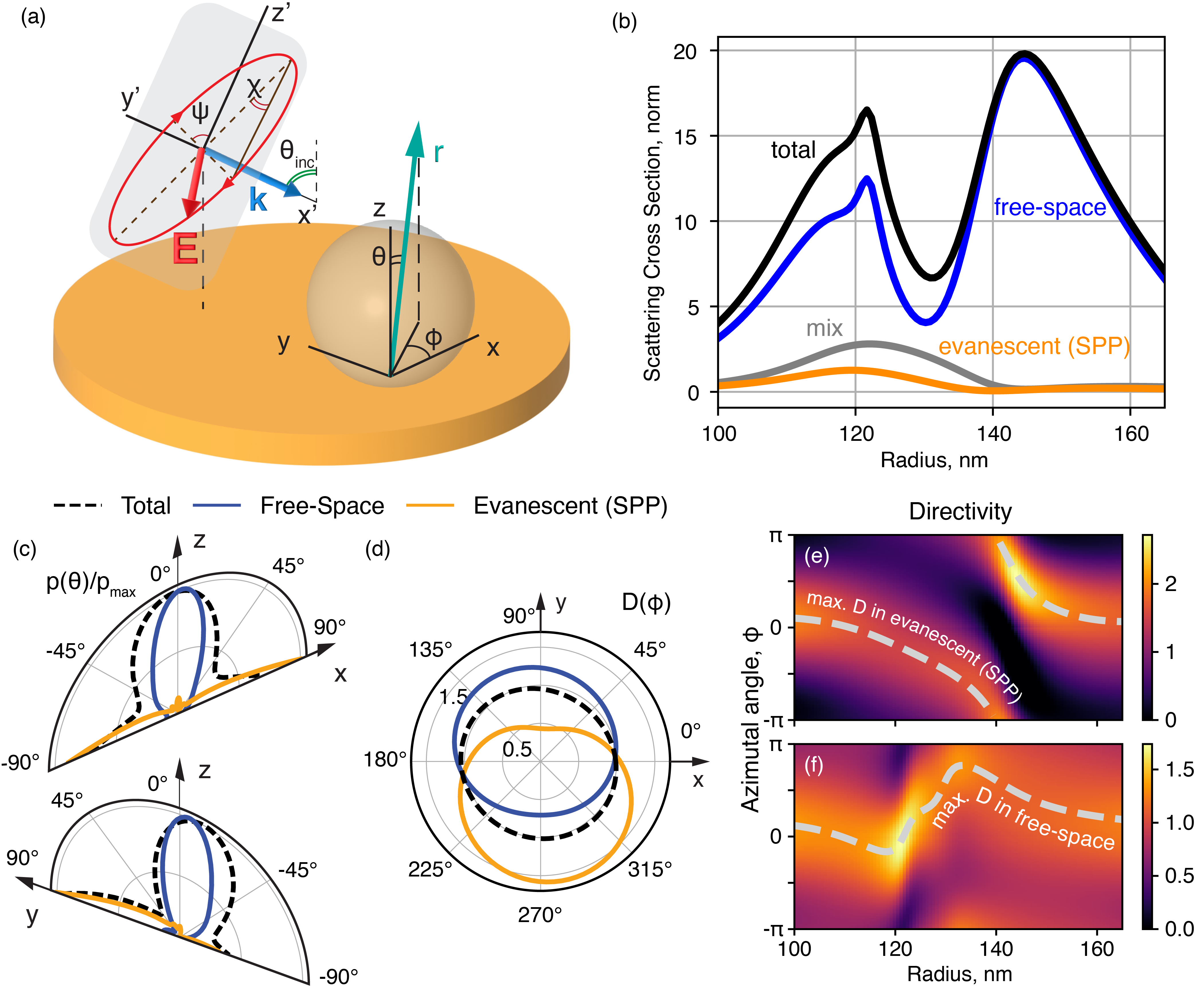}
    \caption{(a) Geometrical configuration of the physical system: a dielectric particle placed above a metal substrate is irradiated with elliptically polarized light. (b) Scattering cross sections into free space, $\sigma_{sc}^{\mathrm{free}}$, into SPP, $\sigma_{sc}^{\mathrm{evan}}$, and their interference, $\sigma_{sc}^{\mathrm{mix}}$. The cross sections are normalized to the geometrical cross section of the particle. For a particle of radius $R = 130$ nm: (c) normalized angular scattering density in the $xz$- and $yz$-planes for the free-space, SPP, and total scattered fields, (d) directivity patterns of the same three contributions. (e) Map of the SPP directivity, $D_{\mathrm{evan}}$, as a function of the in-plane angle $\varphi$ and particle radius $R$. (f) Map of the free-space directivity, $D_{\mathrm{free}}$, as a function of the in-plane angle $\varphi$ and particle radius $R$.}
    \label{fig:figure2}
\end{figure}

\subsection{Directional scattering by a resonant Mie particle}

We consider a resonant dielectric nanoparticle placed at a distance $z_0$ above a metal substrate. 
Throughout this work, the particle response is restricted to the dipole regime, where the electromagnetic interaction is fully described by the induced electric and magnetic dipole moments
\begin{equation}
    \vb{p} = \alpha_E \vb{E}_{\text{loc}}, 
    \qquad
    \vb{m} = \alpha_H \vb{H}_{\text{loc}},
\end{equation}
with $\alpha_E$ and $\alpha_H$ determined by the first Mie scattering coefficients~\cite{Bohren2007}. The local electromagnetic field acting on a particle is decomposed into the incident ($\vb E_0,\vb H_0$) and scattered ($\vb E_{\text{sc}},\vb H_{\text{sc}}$) parts,
\begin{align}
    \vb{E}_{\text{loc}} &= \vb{E}_0 + \vb{E}_{\text{sc}}, \\
    \vb{H}_{\text{loc}} &= \vb{H}_0 + \vb{H}_{\text{sc}}.
\end{align}
In the following, we assume that the particle is excited by a plane wave, and the incident field includes the incident plane wave and the wave  reflected from the interface. For dipoles located near a planar interface, the scattered fields are expressed through the dyadic Green's functions:
\begin{align}
    \vb{E}_{\text{sc}} &= 
    \frac{k^2}{\varepsilon_0}\tens{G}_{\rm E}(\vb{r},\vb{r}_0)\cdot\vb{p}
    + i\omega\mu_0 \left[\nabla\times\tens{G}_{\rm H}(\vb{r},\vb{r}_0)\right]\cdot\vb{m}, \\
    \vb{H}_{\text{sc}} &= 
    k^2 \tens{G}_{\rm H}(\vb{r},\vb{r}_0)\cdot\vb{m}
    - i\omega \left[\nabla\times\tens{G}_{\rm E}(\vb{r},\vb{r}_0)\right]\cdot\vb{p}.
\end{align}

Here $\omega$ is the angular frequency of the incident electromagnetic field. The wavenumber of the incident wave is denoted by $k = \omega/c$, where $c$ is the speed of light in a vacuum. The constants $\varepsilon_0$ and $\mu_0$ represent the vacuum permittivity and permeability. The tensors $\tens{G}_{\rm E}(\vb{r},\vb{r}_0)$ and $\tens{G}_{\rm H}(\vb{r},\vb{r}_0)$ are the electric and magnetic dyadic Green’s functions of the system, accounting for the presence of the planar interface and describing the field response at \(\vb{r}\) due to a source dipole located at \(\vb{r}_0\). The spectral representation of the Green’s tensor for a layered structure involves integration over the in-plane wavevector $k_\rho$~\cite{Novotny_Hecht_2012}. The integral naturally separates into two physically distinct domains: the propagating sector $k_\rho<k$, which contributes to radiation into the upper half-space, and the evanescent sector $k_\rho>k$, which corresponds to the excitation of SPPs ~\cite{Petrov2016, Kostina2019, Kostina2021}. For the radiation patterns, the scattered field is evaluated at observation points $\vb{r}\neq\vb{r}_0$, where it also contains the direct dipole radiation described by the vacuum dyadic Green's function $\tens{G}_0$. Grouping $\tens{G}_0$ with $\tens{G}^{\rm E(H)}_{\text{prop}}$ we decompose the total Green's function as
\begin{equation}
    \tens{G}^{\rm E(H)}(\vb{r},\vb{r}_0)
    = \tens{G}^{\rm E(H)}_{\text{free-space}}(\vb{r},\vb{r}_0)
    + \tens{G}^{\rm E(H)}_{\text{evan}}(\vb{r},\vb{r}_0), \quad
    \tens{G}^{\rm E(H)}_{\text{free-space}} = \tens{G}_0
    + \tens{G}^{\rm E(H)}_{\text{prop}}.
\end{equation}
This separation allows us to decompose the scattered field into two channels, i.e. free-space and evanescent (SPP) channels,
\begin{align}
    \vb{E}_{\text{sc}} &= \vb{E}_{\text{free}} + \vb{E}_{\text{evan}}, \quad \vb{H}_{\text{sc}} = \vb{H}_{\text{free}} + \vb{H}_{\text{evan}},
    \label{eq:em_field_sca_spp}
\end{align}
and correspondingly split the time-averaged Poynting vector as
\begin{equation}
    \vb{S}_{\text{sc}} =
    \vb{S}_{\text{free}} +
    \vb{S}_{\text{evan}} +
    \vb{S}_{\text{mix}},
\end{equation}
where
\begin{align}
    \vb{S}_{\text{free}} = \frac{1}{2}\Re\!\left[\vb{E}_{\text{free}}\times\vb{H}^*_{\text{free}}\right], \
    \vb{S}_{\text{evan}} = \frac{1}{2}\Re\!\left[\vb{E}_{\text{evan}}\times\vb{H}^*_{\text{evan}}\right], \\
    \vb{S}_{\text{mix}} = \frac{1}{2}\Re\!\left[\vb{E}_{\text{free}}\times\vb{H}^*_{\text{evan}}
    + \vb{E}_{\text{evan}}\times\vb{H}^*_{\text{free}}\right].
\end{align}

The geometry of the excitation is shown in Figure~\ref{fig:figure2}(a) where circularly (elliptically) polarized plane wave of wavelength $\lambda$ incidents on a nanoparticle placed on top of the air-gold interface at the angle $\theta_{\text{inc}}$. The polarization state is described by two angles $\psi$ and $\chi$ that parametrize the polarization ellipse as follows: angle $0\leq \psi \leq \pi$ defines the rotation of the major axis of the polarization ellipse, whereas angle $-\pi/4 \leq \chi \leq \pi/4$ specifies the ratio between the semi-axes of the polarization ellipse. For instance, right circular polarization (RCP) is defined at $\psi=0$, $\chi=\pi/4$, while left circular polarization (LCP) is defined by $\psi=0$, $\chi=-\pi/4$.

Figure~\ref{fig:figure2}(b) shows the scattering cross sections into free-space, $\sigma_{\rm sc}^{\rm free}$, into the evanescent (SPP) channel, $\sigma_{\rm sc}^{\rm evan}$, and their interference term, $\sigma^{\rm mix}_{\rm sc}$, normalized to the geometrical cross section $\sigma_0 = \pi R_0^2$. In the considered size interval ($R=100-165$~nm), the particle supports electric- and magnetic-dipole Mie resonances, whose spectral/size evolution governs the redistribution of the scattered power between the free-space (air) and evanescent (SPP) channels. The resulting two-peak structure in the total response reflects the interplay of dipolar resonances in the presence of a metallic substrate; substrate-induced modifications of dipolar lineshapes and the corresponding resonance physics were analyzed in detail in Ref.~\cite{Sinev2016LPR}.

The oblique incidence and circular polarization of the field result in breaking the symmetry of the problem with respect to the $xOz$ and  $yOz$ planes and stimulate strongly directional scattering of the EM energy in both free-space and  SPP channels. The origin of the directional scattering in free space is related to the interference of electric and magnetic dipoles (Kerker effect), while the presence of the substrate adds additional complexity due to metal-induced dipole-images. At the same time, the presence of the circularly polarized electric and magnetic dipoles gives rise to directional excitation of SPPs \cite{Sinev2020ACS}.  The angular signatures of this redistribution are shown in Figure~\ref{fig:figure2}(c,d) for a representative particle size $R=130$~nm.  The angular scattering density is defined as
\begin{equation}
    p_i(\varphi,\theta) =
    \frac{1}{I_0} \vb{S}_i(\varphi,\theta,r\rightarrow\infty)\cdot\vb{n}(\varphi,\theta),
\end{equation}
where $i\in\{\text{free},\text{evan}\}$, $\vb{n}$ is the normal vector. Angular scattering density in Figure~\ref{fig:figure2}(c) is normalized by the maxima of each corresponding line $p_{max}$. 
Since SPPs propagate along the interface, their directivity is evaluated in the plane $\theta=\pi/2$,
\begin{equation}
    D_{\rm evan}(\varphi) =
    \frac{2\pi\, p_{\rm evan}(\varphi,\pi/2)}
    {\int_0^{2\pi} p_{\rm evan}(\tilde{\varphi},\pi/2)\, d\tilde{\varphi}},
\end{equation}
whereas radiation into the free-space is characterized by two angles. However, from the perspective of the generated optical force, we will be interested in the directivity projected on the interface plane that requires integration over the polar angle
\begin{equation}
    D_{\text{free}}(\varphi) =
    \frac{2\pi \int_0^{\pi/2} p_{\text{free}}(\varphi,\theta)\sin\theta\, d\theta}
    {\int_0^{2\pi}\!\int_0^{\pi/2}
    p_{\text{free}}(\tilde{\varphi},\theta)\sin\theta\, d\theta\, d\tilde{\varphi}}.
\end{equation}

Figure~\ref{fig:figure2}(c) presents the normalized angular scattering density in the $xOz$ and $yOz$ planes, comparing the radiation into the free-space (blue), the SPP-associated contribution (orange), and the total scattered field (black dashed). The key observation is that the angular patterns of the air and SPP channels are generally not co-directed: the direction maximizing radiation into the upper half-space differs from the direction maximizing SPP launching along the interface. This mismatch is a direct consequence of electric–magnetic dipole interference in the presence of the interface, where the same induced dipoles couple differently to propagating and evanescent parts of the Green’s function spectrum.

The size dependence of the directional response is summarized in Fig.~\ref{fig:figure2}(e,f). Figure~\ref{fig:figure2}(e) shows a map of the SPP directivity $D_{\rm evan}(\varphi)$ versus azimuthal angle $\varphi$ and particle radius $R$; the dashed curve traces the azimuth $\varphi(D_{\rm evan}^{\rm max})$ corresponding to the direction of maximum SPP launching. In the dipolar regime, changing $R$ effectively tunes the relative weight and phase of the electric and magnetic dipoles, which result in a continuous rotation of the preferential SPP launching direction across a broad angular range. This “full-angle” steering of SPP emission from a single dielectric nanoantenna, achieved through dipole-interference control, has been analyzed and demonstrated in Ref.~\cite{Sinev2020ACS}.

In contrast, Fig.~\ref{fig:figure2}(f) shows the corresponding map for the free-space channel, $D_{\rm free}(\varphi)$, together with the direction of its maximum $\varphi(D_{\rm free}^{\rm max})$. The qualitative difference between panels (e) and (f)—both in the location and in the evolution of the maxima—highlights that SPP launching and far-field radiation are governed by distinct interference conditions, even though they originate from the same induced dipole moments.

\newpage

\begin{figure}[htb!]
    \includegraphics[width=0.39\linewidth]{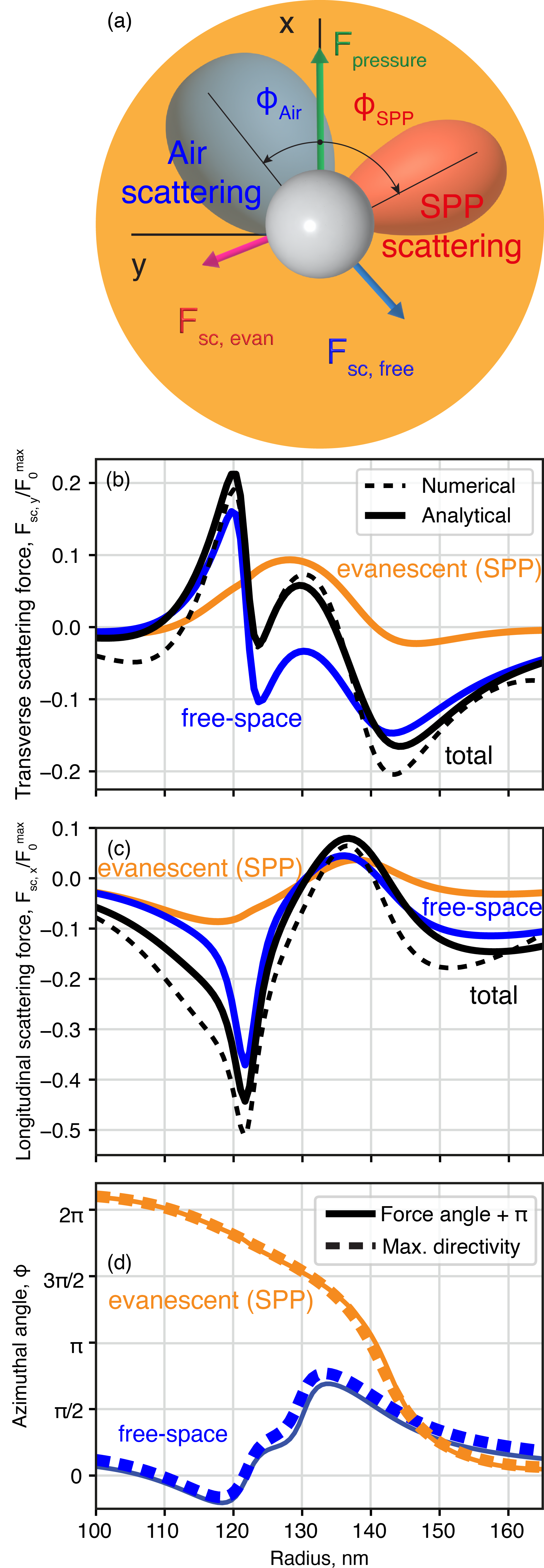}
    \caption{(a) Schematic representation of recoil optical force components $\vb{F}_{\text{sc, free}}$, $\vb{F}_{\text{sc, evan}}$ and angles of maxima SPP launching and free-space scattering in-plane directions. (b) Transverse force versus particle radius: SPP (orange), free-space (blue), and total (black). (c) Longitudinal force versus particle radius with the same color coding. (d) In-plane force angles and corresponding scattering maxima versus particle radius; solid curves denote force directions (blue: free-space, orange: SPP), dashed curves denote the maxima shifted by $\pi$.
    }
    \label{fig:figure3}
\end{figure}

\subsection{Recoil optical force due to free-space and SPP scattering}

Within this section, we analyze the emergence of the recoil optical force due to directional rescattering of the incident field momentum. As mentioned earlier, there are two main scattering channels associated with free-space scattering and SPP generation. One of the main claims is that the direction of the optical force fully corresponds to the direction of free-space and SPP scattering. To demonstrate this, we will decompose the optical force into terms corresponding to SPP generation and free-space scattering.

We start with the well-known expression for the optical force acting on a particle with electric $\vb{p}$ and magnetic $\vb{m}$ dipole response form~\cite{Chen2011}:
\begin{equation}
    \vb{F} = \underbrace{\frac{1}{2} \Re{\vb{p}^* \cdot(\nabla \vb{E}_{\rm loc})}}_{\vb{F}^{\text{e}}} + \underbrace{\frac{1}{2} \mu_0 \Re{\vb{m}^*  \cdot (\nabla \vb{H}_{\rm loc})}}_{\vb{F}^{\text{m}}} \underbrace{- \frac{k^4}{12 \pi c \varepsilon_0} \Re{\vb{p} \times \vb{m}^*}}_{\vb{F}^{recoil}} \;. \label{eq:force_dipole}
\end{equation}
Here, the optical force contains three terms: the force acting on an electric dipole in the electric field $\vb{F}^e$, the force acting on a magnetic dipole in the magnetic field $\vb{F}^{m}$, and the interference (recoil) term $\vb{F}^{\text{recoil}}$ which originates from the directional scattering of the field momentum. In the non-homogeneous environment, the induced dipole moments may have additional contributions due to interaction with the surrounding structure. In our case, the substrate modifies the dipole response due to multiple scattering of the waves, which can be accounted via the effective electric and magnetic polarizability tensors~\cite{Miroshnichenko2015}:  
\begin{align}
    \vb{p} = \hat{\alpha}_{\rm ee} \varepsilon_0 \vb{E}_0 + \hat{\alpha}_{\rm em} \vb{H}_0 \\
    \vb{m} = \hat{\alpha}_{\rm me} \vb{E}_0 + \hat{\alpha}_{\rm mm} \vb{H}_0
\end{align}
where the effective polarizabilities $\hat{\alpha}_{\rm ee}$, $\hat{\alpha}_{\rm em}$, $\hat{\alpha}_{\rm me}$, $\hat{\alpha}_{\rm mm}$  are introduced and characterize the response of particles to the incident (external) fields $\vb E_0, \vb H_0$ as well as their coupling rather than the local fields $\vb E_{\rm loc}, \vb H_{\rm loc}$. The latter ones contains the incident field $\vb E_0$ as well as the scattered field $\vb E_{\text{sc}}=({k^2}/{\varepsilon_0}) \GE (\vb{r}_0)  \cdot \vb{p} + i \omega \mu_0 [\rGH(\vb{r}_0) ] \cdot \vb{m}$. Substituting the local fields acting on the dipole particle:
\begin{align}
    \vb{F}^{e} = \underbrace{\frac{1}{2} \Re{ \vb{p}^* \cdot \nabla \vb{E}_0} }_{\vb{F}^{\rm e}_0} + \underbrace{\frac{1}{2} \Re{\vb{p}^* \cdot \nabla \left( \frac{k^2}{\varepsilon_0} \GE (\vb{r}_0,\vb{r}_0)  \cdot \vb{p} + i \omega \mu_0 [\rGH(\vb{r}_0,\vb{r}_0) ] \cdot \vb{m} \right)}}_{\vb{F}_{\rm sc}^{\rm e}}, \\ 
    \vb{F}^{m} = \underbrace{\frac{\mu_0}{2} \Re{ \vb{m}^* \cdot \nabla \vb{H}_0} }_{\vb{F}^{\rm m}_0} + \underbrace{\frac{\mu_0}{2} \Re{\vb{m}^* \cdot \nabla \left( k^2 \GH(\vb{r}_0,\vb{r}_0)  \cdot \vb{m} - i \omega [\rGE(\vb{r}_0,\vb{r}_0) ] \cdot \vb{p} \right)}}_{\vb{F}_{\rm sc}^{\rm m}},
    \label{eq:force_dipole_components}
\end{align}
we formally split the force into two parts: $\vb F_{0}^{\text{e(m)}}$ related to the incident field and $\vb F_{\text{sc}}^{\text{e(m)}}$ related to the scattered field. The scattered force, in turn, has two terms related to the free space modes and evanescent modes in accordance with the Green's function tensor, since $\GE (\GH) $ can be split into the integration over the free space and over the evanescent waves similarly to Eq.~\eqref{eq:em_field_sca_spp}: 
$$\vb F_{\text{sc} }^{\text{e(m)}}=\vb F_{\text{sc, free}}^{\text{e(m)}} + \vb F_{\text{sc, evan}}^{\text{e(m)}}.$$

Finally, the clear physical picture can be obtained once the corresponding terms will be collected: 
\begin{align}
    &\vb F=\vb F_0 + \vb F_{\text{sc, free}} + \vb F_{\text{sc, evan}}, \\
    &\vb F_0= \vb{F}_{0}^{\rm e}+\vb{F}_0^{\rm m},\\
    &\vb F_{\text{sc, free}}=\vb F_{\text{sc, free}}^{\rm e}+\vb F_{\text{sc, free}}^m+\vb F_{\text{recoil}},\\
    &\vb F_{\text{sc, evan}}=\vb F_{\text{sc, evan}}^{\rm e}+\vb F_{\text{sc, evan}}^{\rm m}.
\end{align}
Their physical meaning can be interpreted quite clearly. The component  $\vb F_0$ corresponds to the optical force that appears due to the presence of the incident field. In the considered geometry, that is the standing wave in the $z$-direction; thus, this force contains a gradient-type force along the $z$-axis and {\it a pressure force} acting mainly along the $x$-axis (direction of incidence); thus, it has only $x$- and $z$-components $\vb F_0=F_{0,x}\vb{\hat{e}}_x+F_{0,z}\vb{\hat{e}}_z$.  The component $\vb F_{\text{sc, free}}$ corresponds to the field rescattered into free-space directly (``recoil" force term) and after reflection over the interface (``sc,free" term). Finally, the evanescent term  $\vb F_{\text{sc, evan}}$ contains the force due to the excitation of evanescent fields with the dominant contribution of the SPP wave. 


To compare the scattering-induced optomechanical response with the radiation patterns discussed above, we now focus on the in-plane (lateral) recoil force and its decomposition into the free-space and SPP contributions. In the discussion below, we deliberately separate the incident-field part $\vb F_0$ from the scattering-related terms in order to (i) directly connect the force to the momentum carried away by the scattered fields and (ii) enable a quantitative comparison with the directivity maxima in Fig.~\ref{fig:figure2}. In particular, the force plotted in Fig.~\ref{fig:figure3} corresponds to
\begin{equation}
    \vb F_{\mathrm{sc}} \equiv \vb F - \vb F_0 = \vb F_{\text{sc, free}} + \vb F_{\text{sc, evan}},
    \label{eq:F_scatt_def}
\end{equation}
i.e., only the scattering-induced force, while the incident-field contribution $\vb F_0$ (which contains the radiation pressure and the standing-wave gradient along $z$) is excluded.

Our analysis shows that directional scattering of the energy and momentum into the free-space and evanescent (SPP) fields defines the direction of the optical force. Figure~\ref{fig:figure3}(a) schematically summarizes these physical mechanisms. The induced dipoles reradiate into two channels: propagating modes in the upper half-space (``air'' scattering) and surface-bound SPP modes. Since both channels are strongly directional for oblique circular/elliptical excitation (Fig.~\ref{fig:figure2}), the particle experiences a lateral recoil force opposing the dominant momentum flux in each channel. 
The radius dependence of the transverse scattering-induced force $F_y$ is shown in Fig.~\ref{fig:figure3}(b). Both analytical (solid curves) and numerical (dashed curves) results are presented, demonstrating good agreement across the full size interval. The total transverse component (black) is formed by the competition between the SPP-related contribution (orange) and the free-space contribution (blue). The SPP term remains of one sign over most of the considered range, reflecting the persistent preferential SPP launching direction, whereas the free-space term exhibits a pronounced dispersive behavior near the dipolar resonances. As a result, the total $F_y$ shows distinct extrema and may change sign when the dominant momentum channel switches from one scattering pathway to the other.

Figure~\ref{fig:figure3}(c) shows the longitudinal scattering-induced component $F_x$ of \eqref{eq:F_scatt_def}.   In contrast to the transverse force $F_y$, the longitudinal response is more sensitive to the redistribution of power between forward/backward propagating modes in the upper half-space and to the SPP momentum component projected onto the $x$ axis. Near the dipolar resonances, where the relative phase between $\vb p$ and $\vb m$ varies rapidly, the longitudinal recoil exhibits large variations in magnitude and may reverse its sign. Importantly, one should note, the $x$-projection of $\vb F_{\mathrm{sc}}$  does not include the radiation pressure exerted by the incident field and represents the recoil associated with scattering only. The incident-field radiation pressure (contained in $\vb F_0$) would add a typically dominant contribution along the direction of the incident wavevector and is intentionally omitted here to preserve a direct connection to the scattering directivities.

The directional correspondence between recoil forces and scattering maxima is demonstrated in Fig.~\ref{fig:figure3}(d). The dashed curves reproduce the azimuthal angles of the directivity maxima for the SPP and free-space channels, $\varphi(D^{\max}_{\text{evan}})$ and $\varphi(D^{\max}_{\mathrm{free}})$, extracted from Fig.~\ref{fig:figure2}(e,f). The solid curves show the azimuthal angles of the corresponding in-plane force vectors, $\varphi(\vb F_{\text{evan (SPP)}})$ and $\varphi(\vb F_{\mathrm{free}})$, shifted by $\pi$ to account for the recoil nature of the force. The close overlap between the dashed and solid curves confirms that the direction of each force component is set by the momentum carried away in the respective scattering channel. Therefore, the same dipole-interference mechanism that enables continuous steering of SPP directivity with particle size (Fig.~\ref{fig:figure2}(e)) simultaneously enables continuous steering of the SPP-related recoil force, while the free-space force follows the evolution of the far-field directivity maximum (Fig.~\ref{fig:figure2}(f)).

Finally, we note that in practical configurations the total force includes $\vb F_0$ in addition to the scattering-induced contributions discussed here. In the present single-beam geometry $\vb F_0$ contains a substantial radiation-pressure component in the plane of incidence and a gradient component along $z$, whereas the transverse component $F_{0,y}$ remains zero by symmetry. This is precisely why the transverse recoil force shown in Fig.~\ref{fig:figure3}(b) provides a particularly direct manifestation of the momentum asymmetry induced by oblique circular/elliptical excitation.

\begin{figure}[th!]
    \centering
    \includegraphics[width=\linewidth]{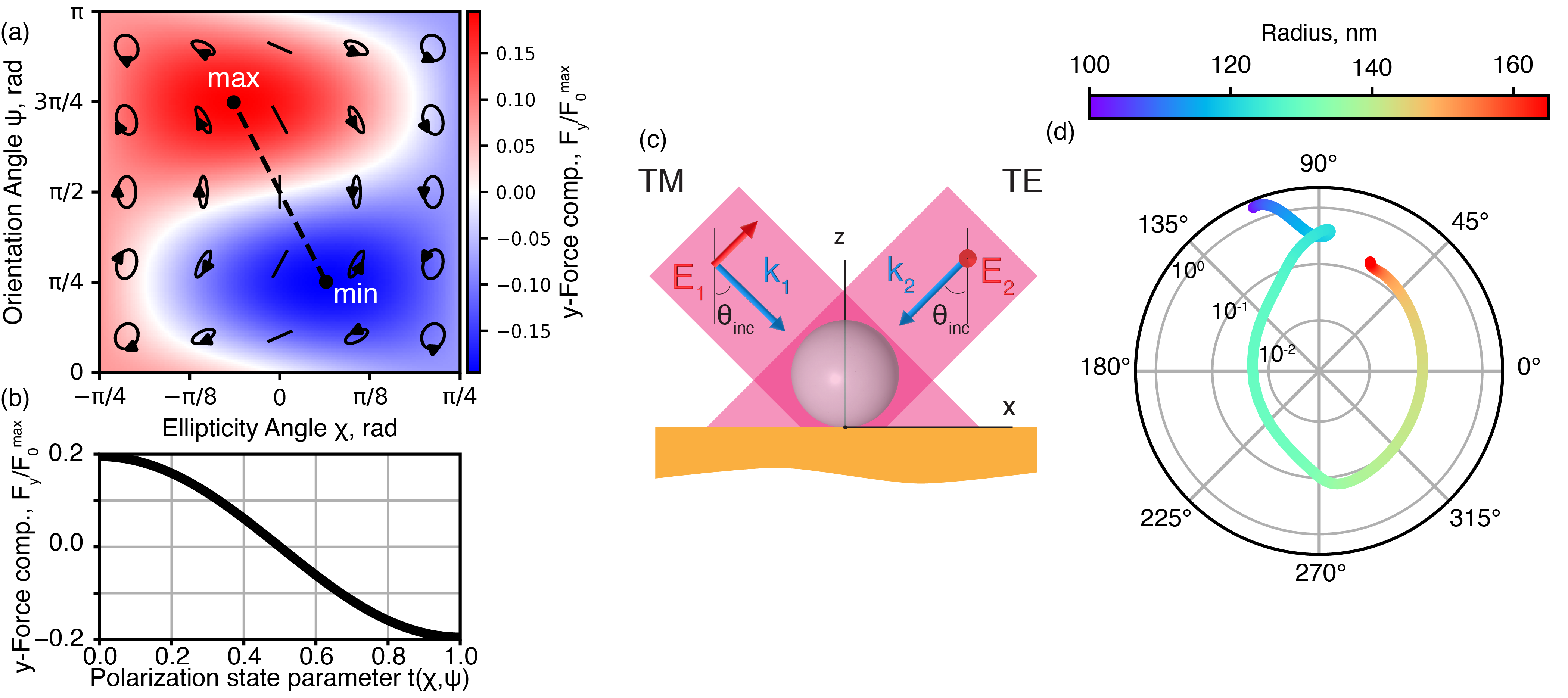}
    \caption{(a) Map of the transverse component of the total optical force $F_y$ as a function of the polarization-ellipse parameters $\psi$ (orientation angle) and $\chi$ (ellipticity angle), for $R = 122$ nm. The inset ellipses show the corresponding polarization states. The black dashed line is a linear path $t = t(\psi, \chi)$ in the $(\psi, \chi)$-plane, running from the minimum to the maximum of the transverse force; (b) Transverse optical force $F_y$ along the parametrization $t = t(\psi, \chi)$ defined in (a), demonstrating continuous control of the transverse optical force by varying the polarization state; (c) Crossed-beam geometry for implementing optical sorting; (d) Direction and amplitude of the in-plane optical force for different particle radii at a fixed polarization state ($\psi = 140^\circ$, $\chi = -5^\circ$).}
    \label{fig:figure4}
\end{figure}

\section{Optical sorting with directional scattering}
The interplay between these two scattering channels allows one to control the direction and amplitude of the in-plane force. While the results shown in Fig.\ref{fig:figure2} and Fig.\ref{fig:figure3} were obtained for a purely RCP wave, one can change the force sign by altering the polarization state of the incident wave. In Fig.~\ref{fig:figure4} (a), a map of the transverse force component is shown as a function of the polarization angles. One can see that by gradually varying the polarization states from right to left polarization with parallel rotation of the ellipse, it appears possible to change the sign and amplitude of $F_{y,\text{sc}}$ component crossing zero at $\psi=\pi/2$ and $\chi=0$ what corresponds to a TE-polarized wave.

The proposed mechanism of the scattered part of the  optical force control and variation can be applied to develop novel methods for the optical sorting of Mie particles by their size. In order to get a higher sensitivity of  the optical force on the particle sizes, one has to suppress the dominant lateral component of the pressure force $\vb F_0$. That  can be achieved in the cross beam geometry, where two beams of orthogonal polarization are incident at opposite angles as shown in Fig.\ref{fig:figure4} (c). Here the $x$-component of the optical pressure force will be compensated by the cross-polarized (TE and TM) beams counter propagating along the interface. The orthogonal polarization prevents the interference of beams along the $x$-axis and compensates the pressure part, so that the scattering part of the force starts to play the dominant role. Optimizing the relative parameters of the beams (relative field amplitudes and phases) allows finding the regime with almost $2\pi$  variation of the lateral components of the optical force for the particle radius in the range from 120 nm to 160 nm making optical sorting of particles by their size possible within the considered geometry. The lateral component of the optical force is shown in Fig.\ref{fig:figure4} (d) for particles of different radius. One can see, that comparing to other approaches \cite{Shilkin2017DirectionalNanoparticles}, here one gets variation of the optical force in almost $2\pi$ range for the considered particle parameters at the fixed parameters of the beams. Further optimization of the beam parameters can provide increased sensitivity by tuning the beam parameters for different particle radii.

At this point, it is important to discuss some expected numerical estimations of the sorting parameter and potential limitations. Firstly, we 
aussume that the particle is illuminated by a laser beam with power $P_{\text{las} } = 100$ mW, focused to a spot of diameter $d_{\text{las}} = 10$ $\mu$m. For such parameters, the intensity of the incident field is on the order of $I \approx 1.3 \times 10^9$ $W/m^2$. The typical optical force $F_0$ acting on the particle due to the incident field is on the order of $F_0 \approx 13$ pN.  If we consider water environment, the typical steady state velocity of the particle is given by $v \approx F_{\rm sc}/(6\pi \eta R) \approx 0.1$ mm/s, where $\eta$ is the viscosity of water. Based on this velocity, we can estimate the time required for the particle to travel the distance of the laser spot $d_{\text{las} }$ as $t \approx d_{\text{las}}/v \approx 0.1$ s. This means that the particle can be sorted by its size within a fraction of a second, which is a promising result for practical applications.

Another important limitation is related to heating. The particle efficiently excite SPP, which produce a strongly localized electromagnetic field at the metal–dielectric interface. This field enhancement leads to increased absorption and, consequently, local overheating of the structure for several tens and even hundreds degrees~\cite{Zhang2021,  milichko_metaldielectric_2017,Zograf2017,Zograf2021}. Moreover, optical heating can results in extemely strong thermooptical nonlinear effects even in single particles \cite{ryabov_nonlinear_2022,nishida_optical_2023, nishida_all-optical_2024}. At the elevated temperatures, stochastic photophoretic forces may arise, with a typical magnitude of $F^{\rm pp} \approx 1$ nN, disturbing controlled optical manipulation~\cite{mi12040466}. However, our simulations show that in the water environment the temperature increase is significantly reduced, $\Delta T \approx 5$ K, due to efficient heat dissipation into the surrounding medium.

\section{CONCLUSION}

In this work, we have demonstrated that optical forces acting on Mie-resonant dielectric nanoparticles near a metal interface are governed by directional scattering into free-space and SPP channels. The interference of electric and magnetic dipole moments leads to strongly asymmetric momentum redistribution, giving rise to a recoil force whose direction follows the scattering directivity in each channel.

By decomposing the force into incident, free-space, and evanescent contributions, we established a direct correspondence between the optical force and the underlying scattering processes. We showed that the competition between these channels enables continuous steering of the in-plane force and its sign reversal in the vicinity of dipolar resonances.

Finally, we proposed a cross-beam configuration that suppresses the radiation-pressure contribution and enhances the role of scattering-induced forces, providing a practical route for optical sorting of resonant nanoparticles by size. Our results highlight the potential of multipolar interference near interfaces as a versatile tool for optical manipulation and force engineering at the nanoscale.

\section{ACKNOWLEDGMENTS}
The work was supported by Russian Science Foundation grant No. 25-22-20034. 
\bibliography{refferences}

\end{document}